# IDStack - The Common Protocol for Document Verification built on Digital Signatures

Chanaka Lakmal[*], Sachithra Dangalla[+], Chandu Herath[§], Chamin Wickramarathna[¶], Gihan Dias[ε], Shantha Fernando[ǂ]
Department of Computer Science and Engineering, University of Moratuwa, Sri Lanka
[*]ldclakmal@gmail.com, [+]sachithradangalla@gmail.com, [§]cbherath93@gmail.com, [¶]chaminbw@gmail.com, [ε]gihan@uom.lk,
[ǂ]shantha@cse.mrt.ac.lk

***Abstract* — The use of physical documents is inconvenient and inefficient in today's world which motivates us to move towards the use of digital documents. Digital documents can solve many problems of inefficiency of data management but proving their authenticity and verifying them is still a problem. This paper presents a solution for this problem using text extraction, digital signatures and a correlation score for a set of documents. The paper discusses the impacts and advantages of the proposed technologies against other possible technologies that could serve the same purpose.**

## I. INTRODUCTION

Use of physical documents is cumbersome, and the world is moving to the use of digital documents. Digital documents are convenient to use but proving their authenticity is often a problem. Since there is no common protocol to validate and verify digitized documents, the process of analyzing the documents in private and government institutions is difficult.

If there is a mechanism to verify the authenticity of documents and to validate the authenticity and content of the documents, it would facilitate users to manage their digital documents. Using a score to determine how much a document is validated and how much a set of documents relate to a user, can be useful. It would provide an alternative support for the user to determine whether to trust the document or not. This research discusses a proposed solution for above mentioned problems. IDStack is a system based on digital signature technology to verify digital documents and provides a confidence score to a document and a correlation score to a set of documents.

This paper is structured as follows. Section II introduces the motivation for the proposed model and Section III discusses the literature of existing technologies that were reviewed and selected and not selected for this system with justifications. Section IV explains the proposed architecture and execution model of the system and Section V discusses future enhancements that could fine-tune this system.

## II. MOTIVATION AND BACKGROUND

In countries like Sri Lanka, the document verification process is still based on printed documents. The document maybe subjected to a process of verification by an authority by using a signature, and this process has proven to be time consuming and cumbersome. Later if there is a need for another verified copy of the same, the whole process will be repeated, or a copy of the document will be verified again by another party. In addition to these difficulties, sometimes the fraudulent activities that are followed by certain citizens lead to doubtfulness of the authenticity of the content of the documents.

This is the problem that is addressed in this paper and a solution architecture and the execution model is proposed which can be used to verify documents incorporating digital signatures and document correlation factors. This will provide a convenient and reliable way of document verification, which will benefit organizations, authorities as well as Sri Lankan citizens.

## III. LITERATURE REVIEW

One of the major issues of the usage of digital documents is the blurred and non-transparent process of verification. Digital documents are more prone to changes, both intentional changes and mistakes. To overcome this issue, people have been trying various methods and schemes. One of the well-known scheme is digital signatures with a public key infrastructure [1]. A valid digital signature ensures three major security requirements for a digital document namely authenticity (document was created by known sender), non-repudiation (sender cannot deny having sent the document), and integrity (document has no get altered in transit).

Nowadays many standalone document viewing and editing applications such as Adobe Acrobat Reader, Microsoft Word and web applications such as DocuSign [2] provide facilities that allow users to incorporate digital signatures in documents.

Another technology that was reviewed for this system is blockchain technology. In the recent past, people have tended to use blockchain technology based schemes for document verification processes. Blockchain concept is mainly used to achieve immutability property using its distributed continuous ledger design. Stampery [3] is one of the implementations which uses Bitcoin [4] and Ethereum [5] blockchains to timestamp and verify data. It uses the OP_RETURN opcode of Bitcoin protocol to embed data by calculating a cryptographic hash of the data. Stampery has stated that it ensures the proof of existence, integrity, ownership, receipt of a given document.

ShoCard is one of the systems developed using blockchain technology for proving a user's identity when dealing with online systems [6]. Each user is given a credit score and based on a defined qualification, (eg credit score>700) institutes can qualify the user identities. It has stated that the users can give banks and other organizations temporary access to the private sections of his blockchain records, to verify his identity. But in ShoCard, as the final security step, a trusted third-party organization such as a financial institution or the government, may need to certify the user's identity which could result in difficulties in practice.

Blocksign is another blockchain based service for legally signing any document, contract, or agreement [7]. It also hashes the document and stores it in a blockchain. But in such a system, verifying a document soon after signing is not possible because adding a transaction to a blockchain needs to go through several stages. To achieve the authenticity, Blocksign stores user account details such as email address along with the cryptographic hash. Integrity is achieved by comparing the hash values in the verifying process. Web applications such as "ProofOfExistence" also work based on the above principle [8].

By reviewing the above implementations, it was observed that a signing and verification system needs either blockchain technology or PKI based digital signatures but not both. Although blockchain provides integrity, to achieve authenticity, the blockchain based document verification solutions need to manage user accounts. But in PKI system the certificate itself has the signer's information. Therefore, it was decided to use PKI in this system.

Another objective of the system is to extract information from a document. The designed protocol in this system extracts information from a pdf document to the json format. Docparser is one of the services which converts pdf documents to machine readable formats such as json, csv, xml etc. [9] The designed protocol extracts information from a pdf document in a similar kind of way. A machine learning process helps the extraction by automatically identifying the fields such as name, address.

## IV. Architecture and Implementation

### A. *Modularized Architecture*

IDStack is a decentralized API stack that can be used by citizens through a client application of their own that can call the IDStack protocol. The protocol is implemented in modularized architecture which has three unique different modules which are focused on three different applications that are connected to the IDStack web service as illustrated in [Fig. 1.]

1. Data extraction module
2. Data validating module
3. Score calculating module

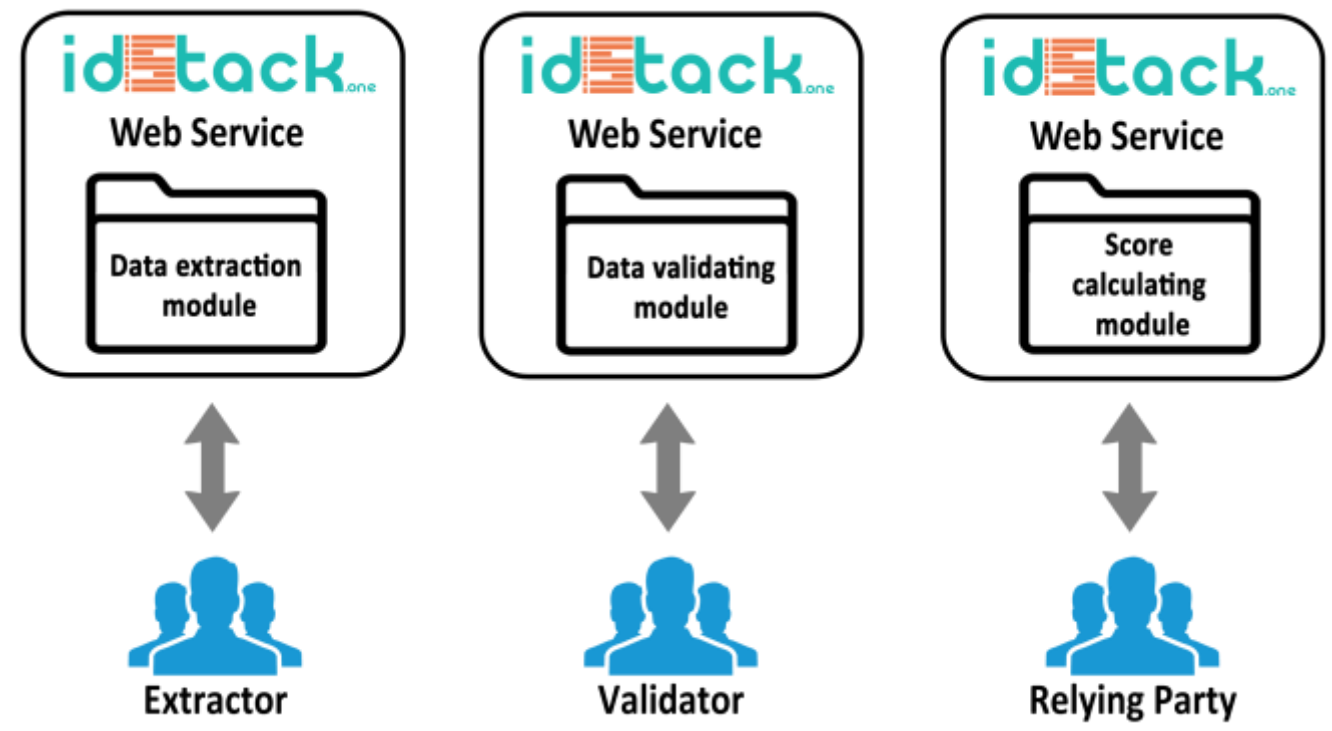

Fig. 1. IDStack modularized architecture

These three different modules are accessed by three different user roles called extractor, validator and relying party. Any citizen can play any role in this protocol. The protocol can be accessed via the IDStack API provided. Every module consists of a set of functionalities which are open to citizens via the API.

Table 1. IDStack user role vs module

| **User role** | **User description** | **Module** | **Module description** |
|---|---|---|---|
| Extractor | The owner or any third-party user can be an extractor that can validate and digitally sign saying that the physical document and the digital document are same.<br>(e.g.: Notary public extracts the data from physical document, creates machine-readable document and digitally signs it) | Data extraction module | This module can create a machine-readable format of a physical/electronic document which is signed by the extractor. |
| Validator | A third-party user that can validate and digitally sign saying that the previous signature is valid or the content is valid or both are valid.<br>(e.g. Vice chancellor of the university says that the signature of senior lecturer is valid) | Data verification module | This module allows user to view the previous digital signatures and add new signatures on the document. |
| Relying Party | The party that has the requirement to view the documents.<br>(e.g.: Foreign embassy views a document or set of documents) | Score calculating module | This module analyses a document or set of documents, evaluates the signatures of documents and correlativity of the documents and calculates a score. |

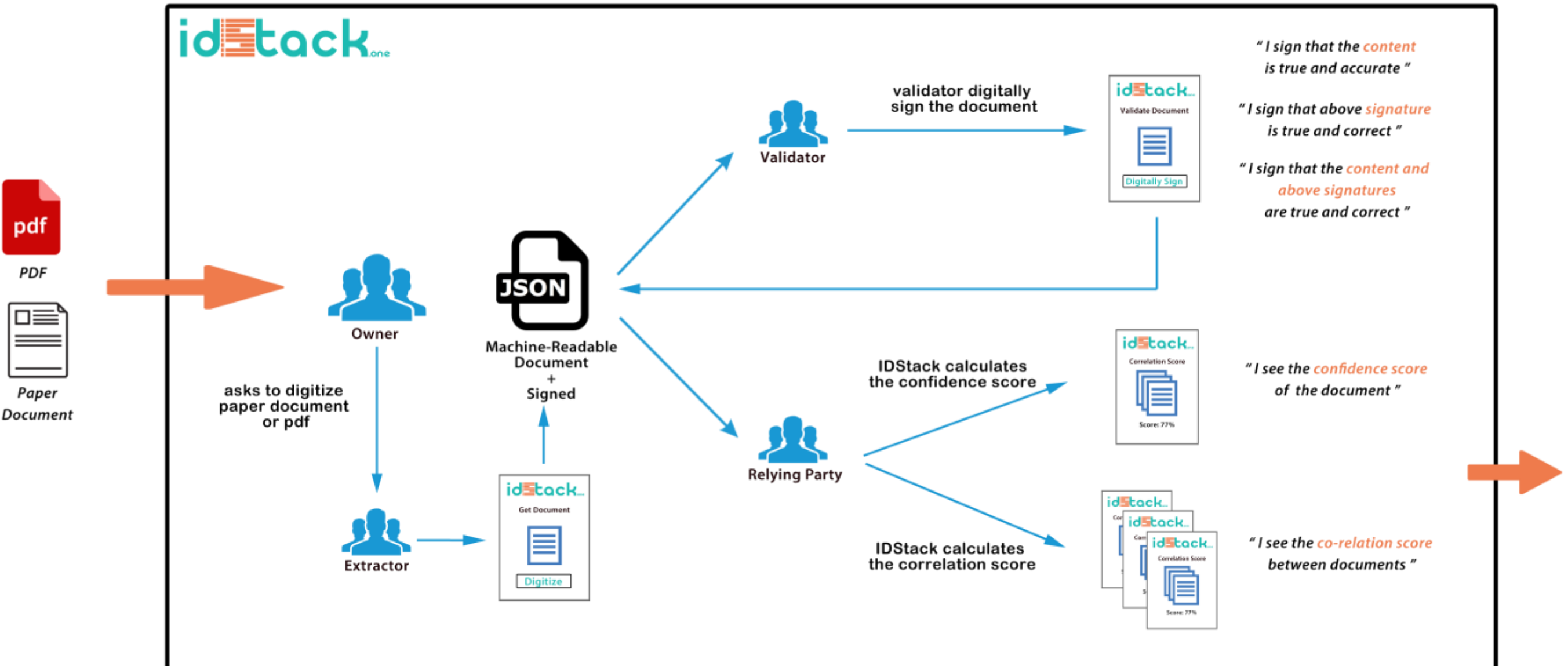

Fig. 2. IDStack protocol

*1) Machine-Readable Document*

The machine-readable document is a JSON file which contains the content of the document in a key value pair design. Apart from that, the digital signature related attributes of the extractor and the set of signers are also included into the same JSON document.

*2) User Roles vs Modules*

The identified roles of users and the respective modules used is presented in [Table 1.]

## B. IDStack Protocol

Citizens who have the documents in digital or physical formats (so called owners of the documents) requests from an extractor to verify and digitize it. The extractor digitizes the document and creates a machine-readable document using IDStack saying that the physical or the digital document and the machine-readable document is 100% matching with each other and then digitally signs with his own signature. The machine-readable document can be signed with any number of validators with their digital signatures. That signature may validate a previous signature or any part of the content of the document or both. Then those signed machine-readable documents can be viewed by the relying party with an automated real-time scoring mechanism which calculates a score based on the signatures and the correlativity of the document parameters. This protocol is illustrated in [Fig. 2.]

# V. Conclusion and Future Work

The need for a better solution for document verification in Sri Lankan context is clear and expanding. With the advancement in the use of digital documents, document verification has a very good potential to grow in many directions. IDStack has proposed and successfully implemented a solution that uses digital signatures, text extraction and document correlation to address the aforementioned issue.

The system is deployed and the API stack is available online [10] and the system has successfully achieved the functional and nonfunctional requirements. The following future contributions are potential gateways that could take the system to the next level.

## A. Language support

Currently, IDStack supports only English documents, which does not satisfy the popular demand of the general Sri Lankan documents. Therefore, the next update would be to support multiple languages, primarily Sinhala and Tamil languages.

## B. Data Extraction

The scope of the current system limits the input documents of the system by allowing only text-based pdf documents. One of the future updates is to support multiple text files and another major update is to support image processing and OCR to extract data from image files.

Another limitation of the system is the use of predefined document templates that maps the extracted data of the input document, to information in machine-readable format. The current system could be further supported by an intelligent system that would automate the template formation or allow users to define their own templates.

## C. Digital signatures

The current system's digital signing is based on self-signed certificates that allow users to create their own signatures. It is a temporary solution that increases the usability of the system, but it imposes the problem of verifying the signer's identity. One solution would be to allow global certificates that are issued by a reputed Certificate Authority which would be costly in the context of the system's target audience. Therefore, an important future work of the system is to find a digital signing mechanism that supports strong identity verification while also being cost-efficient.

### *D. Document Score*

In the current system, the score per document is based on the signatures on the document. An update for this could be to fine-tune the scoring mechanism to consider the document's features such as spellings and semantics to calculate a better confidence score for the document. Furthermore, the score calculation could be accompanied with a validator's comment which would give a human score to the document which would increase the accuracy of the score.

The research done on the correlation score of a set of documents is to be continued to find a better-fitting score to the document's true correlation status using machine learning and information retrieval approaches.